# Proportional Fair Resource Allocation on an Energy Harvesting Downlink - Part I: Structure

Neyre Tekbiyik, Tolga Girici, Elif Uysal-Biyikoglu, and Kemal Leblebicioglu

## Abstract

This paper considers the allocation of time slots in a frame, as well as power and rate to multiple receivers on an energy harvesting downlink. Energy arrival times that will occur within the frame are known at the beginning of the frame. The goal is to optimize throughput in a proportionally fair way, taking into account the inherent differences of channel quality among users. Analysis of structural characteristics of the problem reveals that it can be formulated as a biconvex optimization problem, and that it has multiple optima. Due to the biconvex nature of the problem, a Block Coordinate Descent (BCD) based optimization algorithm that converges to an optimal solution is presented. Numerical and simulation results show that the power-time allocations found by BCD achieve a good balance between total throughput and fairness.

This work was supported by TUBITAK Grant 110E252.

This work has been submitted to the IEEE for possible publication. Copyright may be transferred without notice, after which this version may no longer be accessible.

N. Tekbiyik, E. Uysal-Biyikoglu, and Kemal Leblebicioglu are with the Department of Electrical and Electronics Engineering, Middle East Technical University, Ankara 06800 Turkey. e-mail: ntekbiyik,elif,lebleb@eee.metu.edu.tr

T. Girici is with the Department of Electrical and Electronics Engineering, TOBB Economics and Technology University, Ankara 06560 Turkey. e-mail: tgirici@etu.edu.tr

## I. Introduction

Management of energy consumption is vital for the sustainability of many wireless communication systems. Especially in the past decade, with the advent of ad-hoc wireless networks, energy efficient scheduling policies have been investigated [1], [2], [3]. Due to recent advances in the areas of solar, piezoelectric and thermal energy harvesting, emerging communication devices have been powered by rechargeable batteries which are capable of harvesting energy through solar cells, vibration absorption devices, thermoelectric generators, wind power, etc. Although energy harvesting allows sustainable and environmentally friendly deployment of wireless networks, it requires efficient utilization of time-varying energy. Hence, it shifts the nature of energy-aware solutions from minimizing energy expenditure to optimizing it over time.

It is well known (*e.g.*, [4], [5], [6]) that optimization of a broadcast channel (*e.g.*, the downlink) shared by many users calls for different choices of rate and power allocation to different users depending on the gains, channel conditions, demands of these users, and most importantly, the objective of the optimization. We pose the following problem whose objective is proportional fairness among users: How to allocate among users the transmission power and the proportion of the time between energy harvests, to achieve a good balance between throughput and fairness in an energy harvesting broadcast system. Specifically, we investigate the proportional fairness based utility maximization problem in a time-sharing multi-user additive white Gaussian noise (AWGN) broadcast channel, where the transmitter's battery gets recharged periodically (at known intervals). The transmitter wishes to be fair in delivering data to each receiver during the scheduling period. Energy is assumed to be harvested at the transmitter during the course of transmission. The data, on the other hand, is assumed to be ready at the transmitter before the transmission starts. In general, the arrival processes for the data and the harvested energy can be formulated as stochastic processes [7], [8], requiring an on-line solution that adapts transmission power and rate in real-time. However, as discussed by Yang and Ulukus in [9], this is less tractable. We focus on finding the optimum *off-line* schedule, by assuming that the energy arrival profile at the transmitter is deterministic and known ahead of time in an off-line manner for a time window, called frame, *i.e.*, the energy harvesting times and the corresponding harvested energy amounts are known at the beginning of each frame. The most challenging aspect of this optimization problem is the set of causality constraints introduced by the energy arrival times, *i.e.*, energy may not be used before it is harvested.

There has been considerable recent research effort on optimizing data transmission with an energy harvesting transmitter. In [10], the authors develop a packet scheduling scheme that minimizes the time by which the energy harvesting transmitter delivers all packets to the receiver of a single-user communication system. In [11], the authors extend this work to the multi-user case and, propose an iterative approach that reduces the two-user broadcast problem into a single-user problem as much as possible, and then, utilizes the single-user solution in [10]. [12] treats the time minimization problem for the two-user broadcast channel differently, as it proposes an iterative solution technique by considering two energy arrival slots at a time. These approaches are extended by [13] and [14] to the case of a transmitter with a finite capacity battery. [15] extends [10] one step further to propose the directional water-filling algorithm, which is able to find the optimal energy management schemes for energy harvesting systems operating in fading channels, with finite capacity rechargeable batteries. Both [14] and [15] investigate the following dual offline problems; maximizing the number of bits transmitted with a given deadline constraint, and minimizing the transmission completion time with a given number of bits to transmit.

Unlike the broadcast related studies mentioned above, [9] investigates the dual problems in a multiple access communication system. By using the generalized iterative backward waterfilling algorithm proposed in [9], the authors can simplify the transmission completion time minimization problem into convex optimization problems, and solve the overall problem efficiently. [16] solves the short-term throughput maximization problem for a battery-limited energy harvesting transmitter in a single link topology.

In [17], the authors consider the problem of energy allocation over a finite horizon for point to-point wireless communications, taking into account a time varying channel and energy source, so as to maximize the throughput. They use dynamic programming and convex optimization techniques to obtain the throughput-optimal energy allocation. In [7], Gatzianas et. al. consider an infinite-horizon *online* throughput maximization problem for a rechargeable sensor network. The authors propose a queue stabilizing transmission policy with decoupled admission control and energy allocation to maximize a function of the long term rate achieved per link. Chen et. al. [18] claim that infinite-horizon based solutions can be highly inefficient, especially in the context of networks with energy replenishment. Hence, unlike [7], [18] investigates the finite-horizon throughput maximization problem for a rechargeable sensor network. The main difference between two work is that, in [18] inefficiencies cannot be made to vanish, and thus, new control techniques are needed to mitigate these inefficiencies.

This work differs from the previously mentioned studies particularly in its aim to maximize the throughput in a proportionally fair way, taking into account the inherent differences of channel quality among users. The optimal solution of the problem presented in this paper entails not only the optimal power/rate allocation to users between energy arrivals, but also the optimal time allocation that will maximize the proportionally fair utility function. Due to characteristics of the utility function, the problem presented in this paper can be formulated as a *biconvex* problem and thus it has multiple optima. However, this does not prevent the proposed BCD algorithm to converge to one of the many optima, and thus, achieve optimal utility.

We start by describing the system model in the next section. In Section III, we make the problem statement precise, study the mathematical structure of this problem. The BCD algorithm is described in Section IV, followed by a detailed analysis and discussion of the nature of the solution found by BCD. In Section V, we test the insight gained from analysis about convergence and the nature of the solution, by running the algorithm on numerical examples. We conclude in Section VI with an outline of further directions.

## II. System Model

Consider a time-slotted system where each frame, of length $F_i$, is divided into $K$ slots. There is a single transmitter that transmits to $N$ users by time sharing. The bandwidth for a single link channel is W whereas power spectral density of the background noise is $N_o$. Channel conditions remain constant during the frame (*i.e.*, $g_n$, the gain of user $n$, is chosen to be constant throughout the frame). The transmitter is equipped with a rechargeable battery such that some energy, $E_{ti}$, is harvested from the environment at the beginning of each time slot $t$ of frame $i$. The length of the $t^{th}$ slot of frame $i$ is $T_{ti}$. In the rest, as we are interested in a specific frame, we drop the frame indicator $i$ and define the harvested energy in slot $t$ as $E_t$, and, the length of slot as $T_t$, as shown in Fig. 1. Note that, the slot lengths do not necessarily need to be equal as the energy arrivals may occur in different moments in time. We do not restrict our problem formulation to the case of periodic energy arrivals ( $T_t = T$ for all $t \in \{1, \ldots, K\}$ ). In [19] however, we use periodic energy arrivals assumption to derive the characteristics of the optimal solution of the problem described in this paper, Problem 1. In this offline problem, energy arrival times and amounts that will occur within the frame are known at the beginning of the frame. For a given frame, the transmitter chooses a power level $p_t$ and a time allocation vector $\tau_t = (\tau_{1t}, \ldots, \tau_{Nt})$, for each time slot $t$ of the frame, where $p_{nt} = p_t$ is the selected transmission power for user $n$ during slot $t$ and, $\tau_{nt}$ is the time allocated for transmission to user $n$ during slot $t$.

## III. Problem Statement and Structure

We define the total achievable rate for user $n$ (the total number of bits transmitted to user $n$ within the $i_{th}$ frame), $R_n = \sum_{t=1}^{K} \tau_{nt} W \log_2 \left( 1 + \frac{p_t g_n}{N_o W} \right)$. Our goal is to maximize a total utility, selected as the log-sum of the user rates $\sum_{n=1}^{N} log_2(R_n)$, which is known to result in proportional fairness [20]. However, due to the nature of the time and power allocation problem, and, energy harvesting procedure, some constraints need to be satisfied when maximizing the utility function. Accordingly, we define the following constrained optimization problem, Problem 1, where (1) represent the nonnegativity constraints. The set of equations in (2), called time constraints, ensure that the total time allocated to users does not exceed the slot length. The set of equations in (3), on the other hand, are technical constraints included to ensure that every user gets a non-zero time allocation during the frame. Finally, the set of equations in (4), called energy causality constraints, ensure no energy is transmitted before becoming available.

---

*Problem 1:*

$$\text{Maximize: } U(\overline{\tau}, \overline{p}) = \sum_{n=1}^{N} \log_2 \left( \sum_{t=1}^{K} \tau_{nt} W \log_2 \left( 1 + \frac{g_n p_t}{N_o W} \right) \right)$$

$$\text{subject to: } \tau_{nt} \geq 0 \ , \ p_t \geq 0 \tag{1}$$

$$\sum_{n=1}^{N} \tau_{nt} = T_t \tag{2}$$

$$\sum_{t=1}^{K} \tau_{nt} \geq \epsilon \tag{3}$$

$$\sum_{i=1}^{t} p_i T_i \leq \sum_{i=1}^{t} E_i \tag{4}$$

---

where $t = 1, ..., K$ and $n = 1, ..., N$. Please note that, Problem 1 can be written as a minimization problem in which the function to be minimized is $-U(\overline{\tau}, \overline{p})$. One might hope that this problem has a unique minimum. Unfortunately, (1) is a nonlinear non-convex problem with potentially multiple local minima, some of which are also globally optimum. Thus, we can only expect that by proper choice of the initial value, our algorithm converges to a stationary point that is nearby the true optimum. In order to develop such an algorithm, we first decompose the problem into two parts (power allocation, time allocation) and determine some characteristics that will be useful in understanding the problem structure better. Fortunately, these characteristics lead us to Corollary 1 which establishes that this is a biconvex problem, which we then exploit to determine the most appropriate algorithm for solving it.

### A. *Structure of the Optimal Power Allocation Problem*

In this section, we assume that the time allocation, $\overline{\tau}$, is determined, and try to characterize the structure of the optimal power allocation problem for this $\overline{\tau}$. When the only variables are power variables, Problem 1 reduces to the following constrained optimization problem:

---

*Problem 2:*

$$\text{Maximize: } U(\overline{p}) = \sum_{n=1}^{N} f_n(\overline{p})$$

$$\text{subject to: } p_t \geq 0 \ , \ \sum_{i=1}^{t} p_i T_i \leq \sum_{i=1}^{t} E_i \tag{5}$$

---

where $t = 1, ..., K$ and, $f_n$ is a function of the total number of bits sent to user $n$:

$$f_n(\overline{p}) = \log_2 \left( \sum_{t=1}^{K} \tau_{nt} R_{nt} \right) \tag{6}$$

and $R_{nt}$ represents the rate of link $n$ in the $t^{th}$ slot:

$$R_{nt} = W log_2 \left( 1 + L_n p_t \right) \quad \text{where} \quad L_n = \frac{g_n}{N_o W} \tag{7}$$

Lemma 1 will be useful to get a handle on the characteristics of the problem. Although we claim no originality for the results of the lemma, we provide a proof for completeness.

*Lemma 1:*     i) Let $h_1, \ldots, h_K$ be strictly concave functions of $p_1, \ldots, p_K$ respectively, and, $c_1, \ldots, c_K \geq 0$. Then, $l = \sum_{i=1}^{K} c_i h_i$ is concave. If one of the $c_i$'s is positive ($> 0$), then $l$ is strictly concave.

ii) Increasing concave functions of strictly concave functions are strictly concave.

*Proof:* The proof is provided in Appendix A. ∎

*Theorem 1:* Problem 2 can be formulated as a strictly convex optimization problem. Thus, there exists only one global optimum for a given time allocation.

*Proof:* The proof is provided in Appendix B. ∎

*B. Structure of the Optimal Time Allocation Problem*

In this section, we assume that the power allocation across all slots has been determined. Then, given that the power variables are known constants, we determine the characteristics of the time allocation. So Problem 1 reduces to Problem 3, where the only variables are the time variables:

---

*Problem 3:*

$$\text{Maximize: } U(\overline{\tau}) = \sum_{n=1}^{N} s_n(\overline{\tau})$$

$$\text{subject to: } \tau_{nt} \geq 0 \ , \ \sum_{n=1}^{N} \tau_{nt} = T_t \ , \ \sum_{t=1}^{K} \tau_{nt} \geq \epsilon \qquad (8)$$

---

where $t = 1, ..., K$, $n = 1, ..., N$ and, $s_n$ is a function of the time variables:

$$s_n(\overline{\tau}) = \log_2\left(\sum_{t=1}^{K} \tau_{nt} R_{nt}\right) \qquad (9)$$

and $R_{nt}$'s (defined in Eq. 7) are known constants that represent the rate of link $n$ in the $t^{th}$ slot.

Theorem 2 below records the convexity of Problem 3, and leads to one of the main results of this paper, Corollary 1. The proof of Theorem 2 rests on the observation in Lemma 2. Although we claim no originality for the results of the lemma, we provide the details for completeness.

*Lemma 2:*   i) Let $q_1, \ldots, q_K$ be affine functions of $\tau_{n1}, \ldots, \tau_{nK}$ respectively, and $d_1, \ldots, d_K \geq 0$. Then, $m = \sum_{i=1}^{K} d_i q_i$ is affine.

ii) Increasing concave functions of affine functions are concave.

*Proof:* The proof is provided in Appendix C.                               ∎

*Theorem 2:*  Problem 3 can be formulated as a convex optimization problem. Thus, all local optima are global optima.

*Proof:* The proof is provided in Appendix D.                               ∎

Note that, Problem 3 is convex, but not necessarily *strictly* convex. Therefore, in general, rather than a unique global optimum, there may be multiple local optima which are all also globally optimum.

*Corollary 1:* Problem 1 can be formulated as a biconvex optimization problem.

*Proof:* The proof is provided in Appendix E. ∎

## IV. Solution Method

In the previous section, we have shown that Problem 1 can be formulated as a biconvex optimization problem since $-U(\overline{\tau}, \overline{p})$ is a biconvex function. Such functions are well-studied in the optimization literature [21], [22]. While not convex, they admit efficient coordinate descent algorithms that solve a convex program at each step. In this section, we present a block coordinate descent based algorithm, shortly BCD, for solving Problem 1. In the BCD solution method, sequentially one block of variables is minimized under corresponding constraints while the remaining blocks are fixed. We have the simplest case of only two block variables $\overline{\tau}$ and $\overline{p}$. Hence, the algorithm alternates between minimization with respect to $\overline{\tau}$ and minimization with respect to $\overline{p}$. Our BCD algorithm operates explicitly as follows:

1) Start from any valid time allocation, for example assign each time slot to different user in the form of TDMA. Assuming that all of the energy $E_t$ is used up until the end of period $t$, the power is determined. This power setting satisfies Eq. (4).

2) Keep $\tau_{nt}$ fixed for all n and t. Optimize $U(\overline{\tau}, \overline{p})$ with respect to $p_t, t = 1, \ldots, K$ and the constraints given by (4).

3) Repeat the following for all $t = 1, \ldots, K$.: Keep $\tau_{ni}$ fixed for all $n = 1, \ldots, N$ and $i \neq t$. Also keep $p_t$ fixed for all $t$. Maximize $U(\overline{\tau}, \overline{p})$ with respect to $\tau_{nt}, n = 1, \ldots, N$ and constraint in Eq. (2).

4) If the variables have converged, stop. Otherwise, go to Step 2.

For optimization of the time variables, the Lagrange multiplier method is used. The optimization of the power variables, however, is accomplished by using the Sequential Uncon-strained Minimization Technique (SUMT) [23]. SUMT is an optimization method that converts a constrained optimization problem into an unconstrained one by adding the constraints to the objective function as a "penalty". It then uses a standard unconstrained optimization algorithm (*e.g.*, Newton, Steepest Descent, etc.) [24], [25] to solve the problem with the new objective function.

Regarding the issue of convergence, Problem 1 is a biconvex optimization problem and as such potentially, there exist many local optima. Therefore, convergence to the global optimum is not guaranteed. However, provided that some conditions are satisfied, convergence to a partial optimum (see Definition 1) is guaranteed. As also discussed by Lin in [26], convergence to a stationary (or critical) point, for block coordinate descent methods requires sub-problems to have unique solutions ( [27], [28]), but this property does not hold here: Although sub-problem 2 is strictly convex, 3 is not strictly convex (only convex). Fortunately, for the case of two blocks, Grippo and Sciandrone [29] have shown that this uniqueness condition is not needed. Hence, BCD converges to a stationary point of Problem 1. As a stationary point can be; minimum, maximum, or a saddle point, this convergence result may not be sufficient. However, we can still use the following definition and theorem (Definition 4.1 and Theorem 4.2 of [21], respectively) to build a stronger result. For this, let $X \subseteq \Re^n$ and $Y \subseteq \Re^m$ be two nonempty sets, let $B \subseteq X \times Y$, and, let $B_x$ and $B_y$ denote the x-sections and y-sections of $B$, respectively.

*Definition 1:* Let $f : B \to \Re$ be a given function and let $(x^*, y^*) \in B$. Then, $(x^*, y^*)$ is called a partial optimum of $f$ on $B$, if

$$f(x^*, y^*) \leq f(x, y^*) \quad \forall x \in B_y^* \quad and \quad f(x^*, y^*) \leq f(x^*, y) \quad \forall y \in B_x^* \tag{10}$$

*Theorem 3:* Let $B$ be a biconvex set and let $f : B \to \Re$, be a differentiable, biconvex function. Then, each stationary point of $f$ is a partial optimum.

Furthermore, Theorem 4.9 of [21] shows that, for BCD-like algorithms[1] (There are only two block of variables, and, sequentially one block of variables is minimized under corresponding constraints and the other block is fixed), when one of the subproblems is strictly convex, all partial optima have the same function value. Hence, we conclude that the BCD algorithm surely converges to a partial optimum of Problem 1, and all partial optima of Problem 1 yield the same utility value. Note that although the final allocation, $(\overline{\tau^*}, \overline{p^*})$ generated by the BCD algorithm, might be a partial optimum, it neither has to be a global nor a local optimum to the given biconvex optimization problem. The following theorem by Wendell and Hurter [30] describes the connection between partial and local optima. For the following biconvex minimization problem,

$$min\,\{f(x,y):\ x \in X \subseteq \Re^n,\ y \in Y \subseteq \Re^m\} \tag{11}$$

*Theorem 4:* Let $X \subseteq \Re^n$ and $Y \subseteq \Re^m$ be convex sets and let $f : B \to \Re$ be a biconvex function with a partial optimum in $(x^*, y^*) \in X \times Y$. Also, let $U(y^*)$ be the set of all optimal solutions to the problem in (11) with $y = y^*$ and let $U(x^*)$ denote the set of all optimal solutions to (11) with $x = x^*$. If $(x^*, y^*)$ is a local optimum of (11), then it necessarily holds that

$$f(x^*, y^*) \le f(x, y)\ \ \forall x \in U(x^*)\ \ \forall y \in U(y^*) \tag{12}$$

We have observed through extensive numerical analysis that the above condition is invariably satisfied for the partial optimum found by BCD, which strengthens our belief that BCD achieves optimal or close to optimal utility. However, as also noted in [21], the given local optimality condition is in general not sufficient. Indeed, Wiesemann claims in [31] (p. 92) that, even the verification whether a particular solution to a biconvex problem is locally optimal is $NP$-complete. Gorski et.al. [21], on the other hand, claims that to find the global optimum of a biconvex minimization problem by a BCD-like algorithm (ACS [21]), a multistart version of BCD can be used. But, still, there is no guarantee to find the global optimum within a reasonable amount of time or to be sure that the actual best minimum is the global one. Hence, it seems justified to settle for the modest goal to find a partial optimum in our case.

---

[1]ACS (Alternate Convex Search) algorithm proposed in [21].

## V. Numerical and Simulation Results

In this section, we present the numerical and simulation results related to BCD algorithm. Throughout our simulations we use the folowing setup: $W = 1kHz$, $N_o = 10^{-6}W/Hz$. Unless otherwise stated, all powers are in Watts and all energies are in Joules. For the sake of an example, we suppose that there are five users in the system and 10 energy arrivals in 100 secs (frame length). The arrivals are $\overline{E} = [20, 100, 1, 1, 1, 70, 100, 1, 10, 40]$ joules in the $[1^{st}, 2^{nd}, \ldots, 10^{th}]$ slots respectively. The first user is the strongest one, and, other users are ordered in a such way that the preceding user is twice as strong as the previous one, i.e., path losses of the users are; 25, 28, 31, 34, 37 dB respectively. The starting point of the algorithm is the "Spend What You Get" policy (proposed by Gorlatova et. al. [8]) combined with TDMA time allocation. This policy corresponds to using all energy in the epoch it was harvested in, and will be referred to in the rest as SG+TDMA. We performed simulations both for unequal and equal slot lengths. In our simulations, we use the following sequence of slot sizes for 10 slots; $S_1 = [10, 12, 5, 7, 4, 15, 20, 2, 10, 15]$ and $S_2 = [25, 44, 14, 7, 3, 32, 47, 19, 26, 38]$, for the case of unequal slot lengths, and, $\widetilde{S}_1 = [10, 10, \ldots, 10]$ and $\widetilde{S}_2 = [25.5, 25.5, \ldots, 25.5]$ for the case of equal slot lengths. Note that, $S_1$ and $\widetilde{S}_1$ have the frame length of 100 secs, whereas $S_2$ and $\widetilde{S}_2$ have the frame length of 255 secs.

First, we assume that the frame length is 100 secs, and, we illustrate the power iterations of the BCD algorithm, for $S_1$, in Figure 2. The power convergence of the algorithm for periodic energy arrivals ($\widetilde{S}_1$), however, is illustrated in Figure 2. As observed from the figures, rather than transmitting with full power, saving some energy for the future use is preferred. Another observation is about the fast convergence of the algorithm, i.e., the powers seem to rarely change after just a few iterations. In order to observe the effect of frame length and different slot lengths, in Figure 4, we show how utility improves through the iterations for all of the aforementioned slot length sequences, i.e., $S_1$, $\widetilde{S_1}$, $S_2$, $\widetilde{S_2}$. The fast convergence of the BCD algorithm is more evident in this figure. The optimal schedules (power and time), optimal utility and thus, the utility improvement (when compared to SG+TDMA) obtained by BCD, for all four sequences are presented in Table I.

In some energy harvesting systems, transmitters have supercapacitors that can store the harvested energy and supply in every predetermined time window, allowing the case of periodic energy arrivals. In such a case, if no energy is harvested within a slot, we set the amount of harvested energy to 0 for that slot. As observed from Figure 4 and Table I, periodic energy arrivals assumption does not degrade the system performance, yet increases the utility improvement. Moreover, as we have shown in [19], by using the periodic energy arrivals assumption we can analytically derive the characteristics of the optimal solution of Problem 1 and, develop two heuristics that closely track the performance of BCD algorithm. Hence, from now on, we present results only for the case of periodic energy arrivals in 100 secs, $\widetilde{S_1}$.

Throughput improvement is another important criteria in our problem setup. Hence, we next investigate the throughput improvement of the users for increasing path losses. The results are illustrated in Figure 5. In the figure, the Mean Path Loss, is computed as $\widetilde{L} = \frac{1}{N}\sum_{i=1}^{N} L_i$ where $L_i$ represents the path loss of user $i$. As seen from the figure, with minor decrease in the throughput improvement of the stronger users, the weak users receive much more bits than that they used to receive with SG policy and TDMA. As $\widetilde{L}$ increases, the overall throughput improvement also increases. For instance, when $\widetilde{L} = 31$, User 1, User 3, User 4, and, User 5 enjoy approximately 3 %, 1621 %, 361 %, 80 % throughput improvement respectively, while User 1 suffers only 32 % of loss. Clearly, BCD is a proportionally fair algorithm which tries to maximize the utility by meeting certain demands of every user.

We next analyze the effect of number and amount of energy arrivals. Assume that there are six users in the system with the following path losses: 19, 22, 25, 38, 31 dB. The results obtained for six different energy arrival sequences (two for K=10 and four for K=12) are shown in Table II. The arrival sequences are intentionally chosen similar to each other, so that it would be easier to determine the effect of small changes on the utility improvement. As observed, it is not the number of slots (number of energy arrivals) but the amount of every individual energy harvest that determines the utility improvement. The events like; sudden decrease in energy level or harvesting very small amount of energy for a long time increases the utility improvement obtained by BCD, as SG+TDMA policy may cause the base station to stay idle for a long time because it does not save energy for future use.

Knowing that it is not the number but the nature of harvests that affect the utility improvement, from now on, we set the number of energy arrivals to be 10 (K=10) and the harvests to arrive as in the $2^{nd}$ arrival sequence, i.e. [20, 100, 1, 1, 1, 70, 100, 1, 10, 40] so that we can analyze the effect of number of users to the performance of the BCD algorithm. Keeping the number of harvests and harvest values the same, we perform a series of simulations with different number of users. First, the effects of the optimal power-time allocation pairs on utility, utility improvement, and fairness are investigated. In order to be able to analyze all scenarios, in the next three figures, we use the following setups: a) The strongest user in the system has 13 dB path loss, and, every new user that joins the system has 3 dB more path loss than the previous one. b) The strongest user has 19 dB path loss, and, every new user deviates 3 dB. c) The strongest user has 25 dB path loss, and, every new user deviates 3 dB. Hence, 13 dB, 19 dB, and, 25 dB seen in the figures represent the path loss of the strongest user in the system. Figures 6 and 7 show how utility and percentage improvement in utility, respectively, change with the increasing number of users. As seen from the figures, the solution found by BCD exhibits significant improvement over a SG+TDMA schedule. Between two methods, SG+TDMA is the worst one since even with a few users, utility can be improved. The results show that when case c) is valid, a utility improvement of approximately 20% is possible with BCD.

Although we aim at proportional fairness in this work, it may be interesting to analyse max-min fairness of the BCD algorithm. Jain's index [32], [33] is a well-known measure of fairness. The index $FI$ takes the value of 1 when there is a complete fair allocation.

$$FI = \frac{(\sum_{i=1}^{N} x_i)^2}{N \cdot \sum_{i=1}^{N} x_i^2} \tag{13}$$

For computing $FI$, we use the no. of bits transmitted to the users, $x_i = 2^{U_i}$ for $i = 1, \ldots, N$, where $U_i = \log_2 \left( \sum_{t=1}^{K} \tau_{it} R_{it} \right)$. From Figure 8, it is clear that SG+TDMA is worse than BCD in terms of fairness. Especially for eight users, $FI_{SG+TDMA} = 0.41$ whereas $FI_{BCD} = 0.80$. Although low path losses embrace lower utility improvement, they mainly allow BCD algorithm to be very efficient in terms of fairness, *e.g.*, above 0.8. However as the path loss difference between users increase, completely fair allocations may not be the optimal ones. For instance, when there are eight users in the system the path losses of the users in case c) are; 25, 28, 31, 34, 37, 40, 43, 46 dB, yielding an excessive difference of 23 dB between the weakest and the strongest user. In this case, the algorithm should favor user 1 more than it favors user 8, in order to maximize the utility function, causing a proportionally (instead of purely) fair allocation.

## VI. CONCLUSION

This paper investigated the proportional fair power and time allocation problem in an energy harvesting broadcast system. The paper focuses on finding the optimum *off-line* schedule, by assuming that the energy harvesting times and the corresponding harvested energy amounts are known at the beginning of each frame. In order to determine the optimal schedule, the problem is designed as a proportional fairness based utility maximization problem. Detailed analysis of structural characteristics of the problem has been performed, which revealed that it can be formulated as a biconvex optimization problem, and that it has multiple optima. Furthermore, an algorithm based on block coordinate descent (BCD), that surely converges to a partial optima of the problem, has been showed. Simulation results show that by allocating among users the transmission power and the proportion of the time between energy harvests, BCD achieves a good balance between throughput and fairness on an energy harvesting downlink. BCD algorithm converges to the partial optimal utility and reaches up to 20% utility improvement compared to SG+TDMA schedule.

[19] will investigate the power and time related characteristics of an optimal solution of the proposed problem, and, develop low-complexity heuristic algorithms that will closely track the performance of the BCD algorithm. The analysis, also, will be validated by the consistency of the numerical results and derivations.

## APPENDIX A

## PROOF OF LEMMA 1

i) Let us define $l = \sum_{i=1}^{K} c_i h_i$ where $h_i$ is a strictly concave function of $p_i$, and, $c_i \geq 0$. For $l$ to be concave, it needs to satisy the concavity condition, i.e., $l(\lambda \overline{p_1} + (1-\lambda)\overline{p_2}) \geq \lambda l(\overline{p_1}) + (1-\lambda)l(\overline{p_2})$ where $0 \leq \lambda \leq 1$, for any point $\overline{p_1}, \overline{p_2}$ in the domain of $l$. Thus, proving that $l$ satisfies the concavity condition completes the proof of part (i). We start by

$$l(\lambda \overline{p_1} + (1-\lambda)\overline{p_2}) = \sum_{i=1}^{K} c_i h_i(\lambda p_{1i} + (1-\lambda)p_{2i}) \geq \sum_{i=1}^{K} c_i(\lambda h_i(p_{1i}) + (1-\lambda)h_i(p_{2i})) \quad (14)$$

$$= \lambda \sum_{i=1}^{K} c_i h_i(p_{1i}) + (1-\lambda) \sum_{i=1}^{K} c_i h_i(p_{2i}) = \lambda l(\overline{p_1}) + (1-\lambda)l(\overline{p_2}) \quad (15)$$

where Eq. (14) follows from the strict concavity of $h_i$ function. From the set of equations described above, one can observe that $l$ is a concave function of $\overline{p}$. Note that in Eq. (14), equality may happen only if $c_i = 0$ for all $i$. Hence, if there exists an index $j$ such that $c_i = 0$ for $i = 1, \ldots, j-1, j+1, \ldots, K$ and, $c_i > 0$ for $i = j$, Eq. (14) is satisfied with ">". Then, $l$ is strictly concave in $\overline{p_k}$. This completes the proof of part (i).

ii) This is rather a general result, and hence, due to space constraints, we refer the interested reader to our technical report [34] for the details of our version of the proof.

## Appendix B

### Proof of Theorem 1

Since the constraints of the problem are linear, and, maximizing $U(\overline{p})$ is equivalent to minimizing $-U(\overline{p})$, showing that the utility function, $U(\overline{p})$, is strictly concave will be enough to show that Problem 2 can be formulated as a strictly convex optimization problem. We start by checking the concavity of $R_{nt}$. As $R_{nt}$ is a function of $p_t$, let $h_t = R_{nt}$ (as in the proof of Lemma 1's part $(i)$). The first and second derivatives of $h_t$ are defined as $\frac{\partial h_t}{\partial p_t} = \frac{WL_n/(ln2)}{(1+L_n p_t)}$ and $\frac{\partial^2 h_t}{\partial p_t^2} = \frac{-WL_n^2/(ln2)}{(1+L_n p_t)^2}$ respectively. As, $W$, $L_n$, and $(1+L_n p_t)^2$ are all positive, $\frac{\partial^2 h_t}{\partial p_t^2}$ is definitely negative for all $t = 1, \ldots, K$. Hence, from the second derivative test [23], $h_t$, thus $R_{nt}$, is strictly concave in $p_t$. Furthermore, let $l_n = \sum_{t=1}^{K} \tau_{nt} R_{nt}$ for an arbitrary user $n$. As all $\tau_{nt}$'s (for $t = 1, \ldots, K$) are nonnegative, and, at least one $\tau_{nt}$ is positive for user $n$, from part $(i)$ of Lemma 1, $l_n$ is a strictly concave function of $\overline{p}$. Note that this is true for all users, i.e., $l_n$ is strictly concave for all $n = 1, \ldots, N$. Thus, from part $(ii)$ of Lemma 1, $f_n$ is strictly concave in $\overline{p}$ for all $n = 1, \ldots, N$. The rest of the proof is straight-forward, since the utility function, $U(\overline{p})$, is a positive linear combination of $f_n$'s and thus (from part $(i)$ of Lemma 1) is strictly concave in $\overline{p}$. Hence, the proof is complete.

## Appendix C

### Proof of Lemma 2

i) Let us define $m = \sum_{i=1}^{K} d_i q_i$ where $q_i$ is an affine function of $\tau_{ni}$, and, $d_i \geq 0$. For $m$ to be affine, it needs to satisfy the affinity condition, i.e., $m(\lambda \overline{\tau_1} + (1 - \lambda)\overline{\tau_2}) = \lambda m(\overline{\tau_1}) + (1 - \lambda)m(\overline{\tau_2})$, where $0 \leq \lambda \leq 1$ and $\overline{\tau_k} = [\tau_{k,n1} \ \ldots \ \tau_{k,nK}]^T$ is a point on $m$. Thus, proving that $m$ satisfies the condition completes the proof of part (i). We start by

$$m(\lambda\overline{\tau_1} + (1-\lambda)\overline{\tau_2}) = \sum_{i=1}^{K} d_i q_i(\lambda\tau_{1i} + (1-\lambda)\tau_{2i}) = \sum_{i=1}^{K} d_i(\lambda q_i(\tau_{1i}) + (1-\lambda)q_i(\tau_{2i})) \quad (16)$$

$$= \lambda\sum_{i=1}^{K} d_i q_i(\tau_{1i}) + (1-\lambda)\sum_{i=1}^{K} d_i q_i(\tau_{2i}) = \lambda m(\overline{\tau_1}) + (1-\lambda)m(\overline{\tau_2}) \quad (17)$$

where $\tau_{1i}$, $\tau_{2i}$ are the $i^{th}$ entries of the $\overline{\tau_1}$ vector and $\overline{\tau_2}$ vector respectively, and, Eq. (16) follows from the "affine" property of the $m_i$ function. From Eq. (17), one can clearly observe that $m$ function is an affine function of time variables. Hence, any nonnegative linear combination of affine functions is affine, and, the proof of part $(i)$ is complete.

ii) Similar to part (ii) of Lemma 1, this is a general result that can be found in the literature. Hence, due to space constraints, we refer the interested reader to [34] for the details.

## Appendix D

### Proof of Theorem 2

Similar to the proof of Theorem 1, showing that the utility function, $U(\overline{\tau})$, is concave will be enough to show that Problem 3 can be formulated as a convex optimization problem. We start by checking the concavity of $s_n$ for $n = 1, \ldots, N$. As $R_{nt}$ is a known constant for every $n \in \{1, \ldots, N\}$ and $t \in \{1, \ldots, K\}$, let $d_t = R_{nt}$, and, $q_t = \tau_{nt}$ for an arbitrary user $n$ (as in the proof of Lemma 2's part $(i)$). It is well-known that a linear function is an affine function. As $q_t$ is a linear function of $\tau_{nt}$ for any user $n$, $q_t$ is affine. Now, let $m_n = \sum_{t=1}^{K} \tau_{nt} R_{nt}$ for an arbitrary user $n$. Note that all $R_{nt}$'s are nonnegative constants known apriori. Thus, from part $(ii)$ of Lemma 2, $s_n$ is concave in $\overline{\tau}$ for all $n = 1, \ldots, N$. The rest of the proof is straight-forward, since the utility function, $U(\overline{\tau})$, is a nonnegative linear combination of $s_n$'s and thus (from part $(i)$ of Lemma 1) is concave in $\overline{\tau}$. Hence, the proof is complete.

## Appendix E

## Proof of Corollary 1

A function $f : X \times Y \to \Re$ is called biconvex if $f(x, y)$ is convex in $y$ for fixed $x \in X$ and is convex in $x$ for fixed $y \in Y$ [22]. Since the constraints of the problem are linear, showing that $-U(\overline{\tau}, \overline{p})$ is biconvex will be enough to show that Problem 1 can be formulated as a biconvex optimization problem. $-U(\overline{\tau}, \overline{p})$ is a function of two set of variables; $\overline{\tau}$ and $\overline{p}$. From Theorem 1, given $\overline{\tau}$, $-U(\overline{p})$ is convex. Similarly, from Theorem 2, given $\overline{p}$, $-U(\overline{\tau})$ is convex. Hence, $-U(\overline{\tau}, \overline{p})$ is biconvex, which completes the proof.

## References

[1] Uysal-Biyikoglu E., Prabhakar B., El Gamal A.: Energy-efficient packet transmission over a wireless link. IEEE/ACM Transactions on Networking. vol. 10, 487-499. (2002)

[2] El Gamal A., Nair C., Prabhakar B., Uysal-Biyikoglu E., Zahedi S.: Energy-efficient scheduling of packet transmissions over wireless networks. In: IEEE INFOCOM, pp. 1773-1782. (2002)

[3] Uysal-Biyikoglu E., El Gamal A.: On adaptive transmission for energy efficiency in wireless data networks. IEEE Transactions on Information Theory. vol. 50, 3081-3094. (2004)

[4] Knopp R., Humblet P. A.: Information capacity and control in single-cell multiuser communications. In: Proceedings of IEEE International Communications Conference (ICC), pp. 331-335. (1995)

[5] Tse D. N. C.: Optimal power allocation over parallel Gaussian broadcast channels. In: IEEE ISIT, pp. 27. (1997)

[6] Liu X., Chong E. K. P., Shroff N. B.: Opportunistic transmission scheduling with resource-sharing constraints in wireless networks. IEEE Journal on Selected Areas in Communications. vol. 19, no.10, 2053–2064. (2001)

[7] M. Gatzianas M., Georgiadis L., Control of wireless networks with rechargeable batteries. IEEE Transactions on Wireless Communications. vol. 9, no.2, 581–593. (2010)

[8] Gorlatova M., Berstein A., Zussman G.: Performance Evaluation of Resource Allocation Policies for Energy Harvesting Devices. In: IEEE Symposium on Modeling and Optimization in Mobile, Ad Hoc, and Wireless Networks (WiOpt), pp. 189–196. (2011)

[9] Yang J., Ulukus S.: Optimal Packet Scheduling in a Multiple Access Channel with Energy Harvesting Transmitters. Journal of Communications and Networks, special issue on Energy Harvesting in Wireless Networks, (to appear, 2012)

[10] Yang J., Ulukus S.: Transmission completion time minimization in an energy harvesting system. In: 44th Annual Conference on Information Sciences and Systems (CISS), pp. 1–6. (2010)

[11] Yang J., Ozel O., Ulukus S.: Broadcasting with an Energy Harvesting Rechargeable Transmitter. IEEE Transactions on Wireless Communications. vol. 11, no. 2, 571–583. (2012)

[12] Antepli M. A., Uysal-Biyikoglu E., Erkal H.: Optimal Packet Scheduling on an Energy Harvesting Broadcast Link. IEEE Journal on Selected Areas in Communication. vol. 29, no. 8, 1712–1731. (2011)

[13] Ozel O., Yang J., Ulukus S.: Broadcasting with a battery limited energy harvesting rechargeable transmitter. In: IEEE WiOpt, pp. 205–212. (2011)

[14] Tutuncuoglu K., Yener A.: Optimum transmission policies for battery limited energy harvesting systems. IEEE Transactions on Wireless Communications (submitted). (2011)

[15] Ozel O., Tutuncuoglu K., Yang J., Ulukus S., Yener A.: Resource management for fading wireless channels with energy harvesting nodes. In: IEEE INFOCOM, pp. 456–460. (2011)

[16] Tutuncuoglu K., Yener A.: Short-Term Throughput Maximization for Battery Limited Energy Harvesting Nodes. In: IEEE ICC, pp. 1–5. (2011)

[17] Ho C., Zhang R.: Optimal energy allocation for wireless communications powered by energy harvesters. In: IEEE ISIT, pp. 2368–2372. (2010)

[18] Chen S., Sinha P., Shroff N. B., Joo C.: Finite-Horizon Energy Allocation and Routing Scheme in Rechargeable Sensor Networks. In: IEEE INFOCOM, pp. 2273-2281. (2011)

[19] Tekbiyik N., Uysal-Biyikoglu E., Girici T., Leblebicioglu K.: Proportional Fair Resource Allocation on an Energy Harvesting Downlink - Part II: Algorithms. (Submitted, April 2012).

[20] Mao Z., Koksal C. E., Shroff N. B.: Resource Allocation in Sensor Networks with Renewable Energy. In: IEEE ICCCN, pp. 1–6. (2010)

[21] Gorski J., Pfeuffer F., Klamroth K.: Biconvex sets and optimization with biconvex functions: a survey and extensions. Mathematical Methods of Operations Research. vol. 66, no. 3, 373–407. (2007)

[22] Pirsiavash H., Ramanan D., Fowlkes C.: Bilinear classifiers for visual recognition. IEEE Computer Vision and Pattern Recognition (CVPR). 1482-1490. (2010)

[23] Bazaraa M. S., Sherali H. D., Shetty C. M.: Nonlinear Programming Theory and Algorithms. John Wiley and Sons. (2006)

[24] Boyd S. P., Vandenberghe L.: Convex Optimization. Cambridge University Press. (2004)

[25] Luenberger D. G., Ye Y.: Linear and Nonlinear Programming. Springer Science. (2008)

[26] Lin C.-J.: Projected Gradient Methods for Nonnegative Matrix Factorization. Neural Comput. vol. 19, no. 10, 2756–2779. (2007)

[27] Powell M. J. D.: On search directions for minimization. Mathematical Programming. vol. 4, 193–201. (1973)

[28] Bertsekas D. P.: Nonlinear Programming. Athena Scientific. (1999)

[29] Grippo L., Sciandrone M.: On the convergence of the block nonlinear Gauss-Seidel method under convex constraints. Operations Research Letters. vol. 26, 127–136. (2000)

[30] Wendell R. E., Hurter A. P.: Minimization of a non-separable objective function subject to disjoint constraints. Operations Research. vol. 24, no. 4, 643–657. (1976)

[31] Wiesemann W.: Optimization Of Temporal Networks Under Uncertainty. Springer Heidelberg Dordrecht. (2012)

[32] Jain R., Chiu D., Hawe. W.: A Quantitative Measure Of Fairness And Discrimination For Resource Allocation In Shared Computer Systems. Technical Report. (1984)

[33] Mahmoodi T., Friderikos V., Holland O., Aghvami H.: Balancing Sum Rate and TCP Throughput in OFDMA Based Wireless Networks. In: IEEE ICC, pp. 1–6. (2010)

[34] Tekbiyik N., Girici T., Uysal-Biyikoglu E., Leblebicioglu K.: Proportional Fair Downlink Time Sharing and Power Allocation in the Presence of Energy Harvesting. Technical Report. Middle East Technical University (2012).

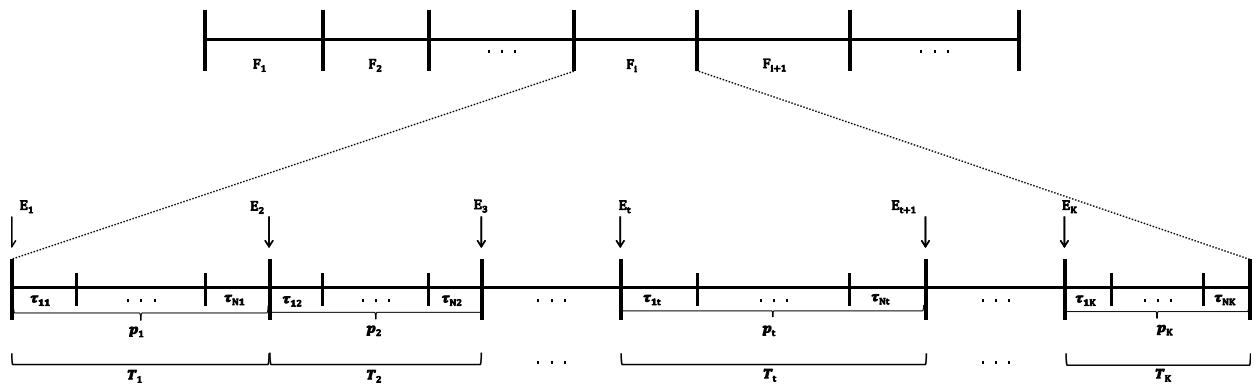

Fig. 1.  Problem illustration: There are $K$ energy arivals in a frame, and, the time between consecutive arrivals are allocated to $N$ users.

TABLE I

THE RESULTS OF BCD ALGORITHM FOR FOUR DIFFERENT SLOT LENGTH SEQUENCES.

| | | Slot 1 | Slot 2 | Slot 3 | Slot 4 | Slot 5 | Slot 6 | Slot 7 | Slot 8 | Slot 9 | Slot 10 | Utility | Utility Imp. |
|---|---|---|---|---|---|---|---|---|---|---|---|---|---|
| **Time Allocation** | Users vs. Slot Lengths | 10 | 12 | 5 | 7 | 4 | 15 | 20 | 2 | 10 | 15 | | |
| | 1 | 10 | 12 | 0 | 0 | 0 | 0 | 3.1288 | 0 | 0 | 0 | | |
| | 2 | 0 | 0 | 0 | 0 | 0 | 5.7638 | 16.8712 | 0 | 0 | 0 | | |
| | 3 | 0 | 0 | 0 | 7 | 4 | 9.2362 | 0 | 0 | 0.1987 | 0 | 75.7273 | % 8.5449 |
| | 4 | 0 | 0 | 0 | 0 | 0 | 0 | 0 | 0 | 9.8013 | 6.9208 | | |
| | 5 | 0 | 0 | 5 | 0 | 0 | 0 | 0 | 2 | 0 | 8.0792 | | |
| | Power Allocation | 2 | 2.0910 | 5.9724 | 3.4337 | 3.4337 | 3.1027 | 2.5535 | 5.9723 | 4.4268 | 5.1636 | | |
| **Time Allocation** | Users vs. Slot Lengths | 10 | 10 | 10 | 10 | 10 | 10 | 10 | 10 | 10 | 10 | | |
| | 1 | 10 | 10 | 6.2337 | 0 | 0 | 0 | 0 | 0 | 0 | 0 | | |
| | 2 | 0 | 0 | 3.7663 | 10 | 10 | 0 | 0 | 0 | 0 | 0 | | |
| | 3 | 0 | 0 | 0 | 0 | 0 | 10 | 10 | 0 | 0 | 0 | 75.7325 | % 9.6133 |
| | 4 | 0 | 0 | 0 | 0 | 0 | 0 | 0 | 10 | 6.3094 | 0 | | |
| | 5 | 0 | 0 | 0 | 0 | 0 | 0 | 0 | 0 | 3.9606 | 10 | | |
| | Power Allocation | 2 | 2.0182 | 2.2189 | 2.5923 | 2.5923 | 3.4327 | 3.4327 | 4.6482 | 5.1876 | 6.2772 | | |
| **Time Allocation** | Users vs. Slot Lengths | 25 | 44 | 14 | 7 | 3 | 32 | 47 | 19 | 26 | 38 | | |
| | 1 | 25 | 44 | 0 | 0 | 0 | 0 | 0.9619 | 0 | 0 | 0 | | |
| | 2 | 0 | 0 | 0 | 0 | 0 | 13.4168 | 46.0381 | 0 | 0 | 0 | | |
| | 3 | 0 | 0 | 14 | 0 | 0 | 18.5832 | 0 | 19 | 0 | 0 | 78.2339 | % 9.0566 |
| | 4 | 0 | 0 | 0 | 0 | 3 | 0 | 0 | 0 | 0 | 38 | | |
| | 5 | 0 | 0 | 0 | 7 | 0 | 0 | 0 | 0 | 26 | 0 | | |
| | Power Allocation | 0.7956 | 0.7956 | 1.3866 | 2.4499 | 1.8390 | 1.2129 | 1.0275 | 1.3866 | 2.4499 | 1.8390 | | |
| **Time Allocation** | Users vs. Slot Lengths | 25.5 | 25.5 | 25.5 | 25.5 | 25.5 | 25.5 | 25.5 | 25.5 | 25.5 | 25.5 | | |
| | 1 | 25.5 | 20.8998 | 0 | 0 | 25.5 | 0 | 0 | 0 | 0 | 0 | | |
| | 2 | 0 | 4.6002 | 25.5 | 25.5 | 0 | 4.2309 | 0 | 0 | 0 | 0 | | |
| | 3 | 0 | 0 | 0 | 0 | 0 | 21.2692 | 25.5 | 3.4418 | 0 | 0 | 78.2314 | % 9.9830 |
| | 4 | 0 | 0 | 0 | 0 | 0 | 0 | 0 | 22.0582 | 18.5955 | 0 | | |
| | 5 | 0 | 0 | 0 | 0 | 0 | 0 | 0 | 0 | 6.9045 | 25.5 | | |
| | Power Allocation | 0.7758 | 0.8157 | 1.0472 | 1.0472 | 0.7758 | 1.3140 | 1.3841 | 1.7590 | 1.9852 | 2.5862 | | |

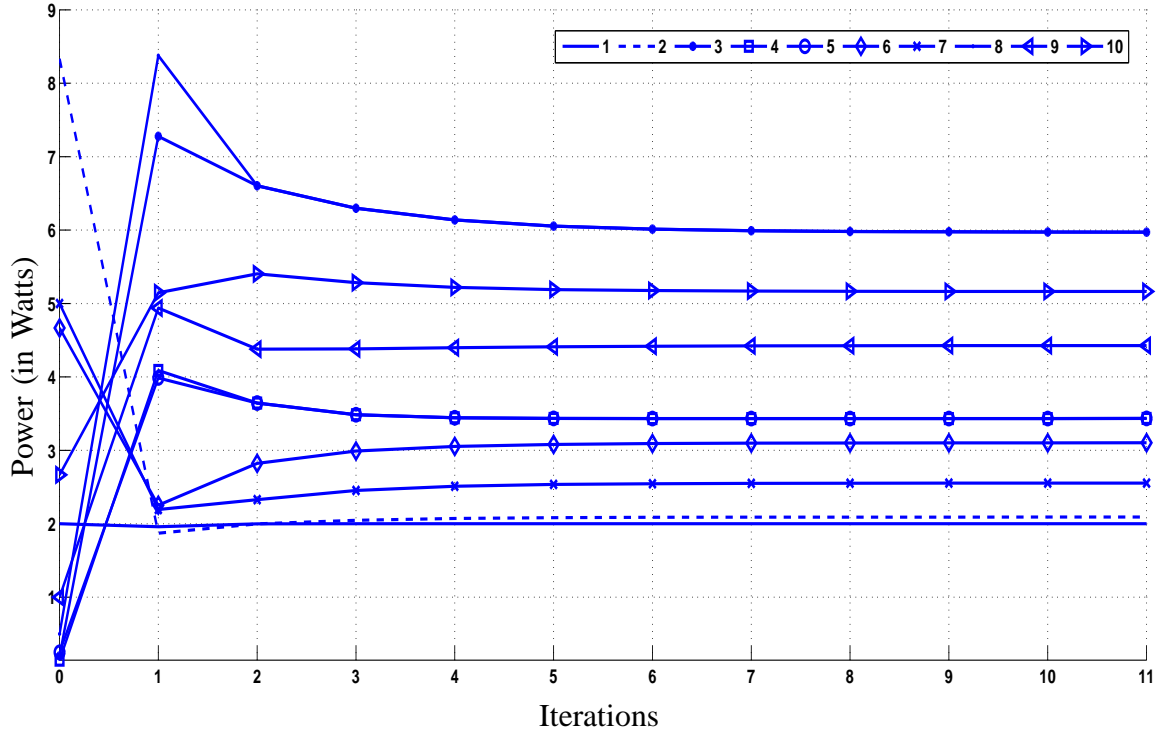

Fig. 2. Powers vs. Iterations (N=5, K=10, Unequal slot lengths): The numbers in the legend represent the corresponding slots. Starting from SG policy, BCD converges to the optimal powers in 11 iterations.

g replacements

TABLE II
THE EFFECT OF NUMBER AND AMOUNT OF ENERGY HARVESTS

| Number of Harvests | Harvests | | | | | | | | | | | | Utility Improvement (%) |
|---|---|---|---|---|---|---|---|---|---|---|---|---|---|
| 10 | Slot 1 | Slot 2 | Slot 3 | Slot 4 | Slot 5 | Slot 6 | Slot 7 | Slot 8 | Slot 9 | Slot 10 | | | |
| | 20 | 100 | 10 | 60 | 10 | 70 | 100 | 10 | 10 | 40 | | | 2.7111 |
| | 20 | 100 | 1 | 1 | 1 | 70 | 100 | 1 | 10 | 40 | | | 7.8467 |
| 12 | Slot 1 | Slot 2 | Slot 3 | Slot 4 | Slot 5 | Slot 6 | Slot 7 | Slot 8 | Slot 9 | Slot 10 | Slot 11 | Slot 12 | |
| | 20 | 60 | 100 | 1 | 1 | 1 | 70 | 85 | 100 | 1 | 10 | 40 | 7.2562 |
| | 20 | 60 | 100 | 0.5 | 1 | 0.5 | 70 | 85 | 100 | 0.5 | 10 | 40 | 8.4308 |
| | 20 | 60 | 100 | 0.5 | 50 | 0.5 | 70 | 85 | 100 | 0.5 | 10 | 40 | 6.7132 |
| | 20 | 60 | 100 | 1 | 0.5 | 0.5 | 1 | 1 | 100 | 0.5 | 10 | 40 | 8.5150 |

g replacements

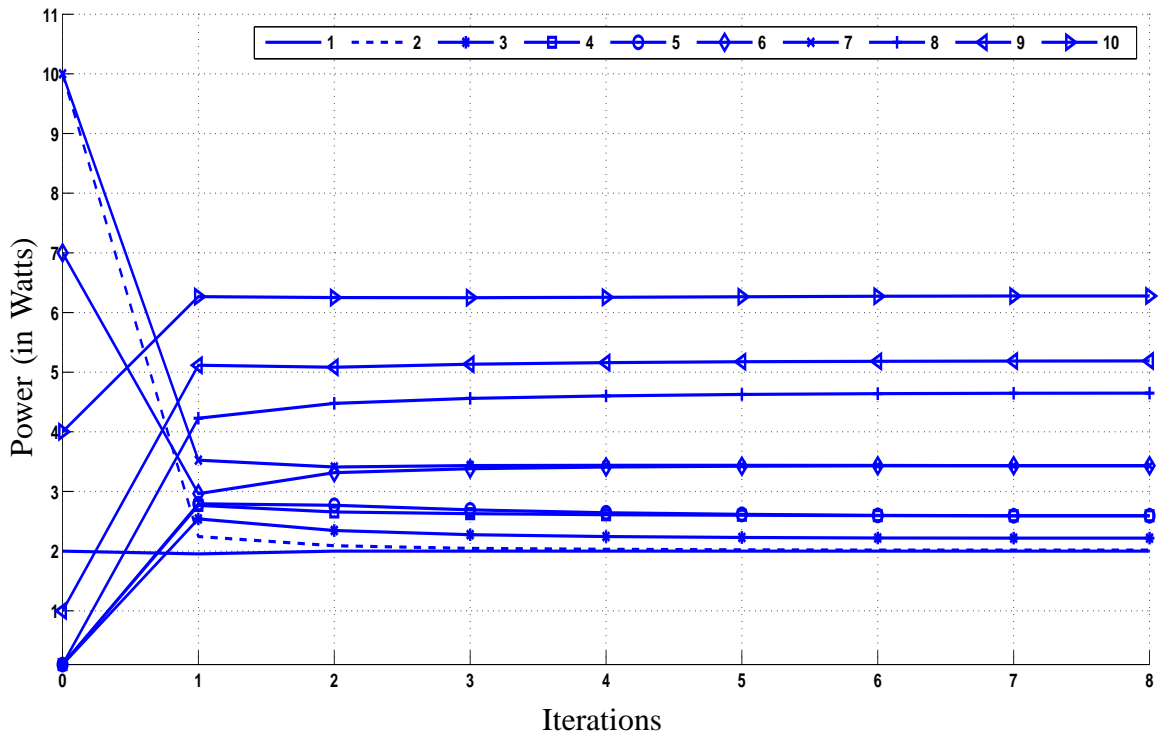

Fig. 3.   Powers vs. Iterations (N=5, K=10, Equal slot lengths): The numbers in the legend represent the corresponding slots. Starting from SG policy, BCD converges to the optimal powers in 8 iterations.

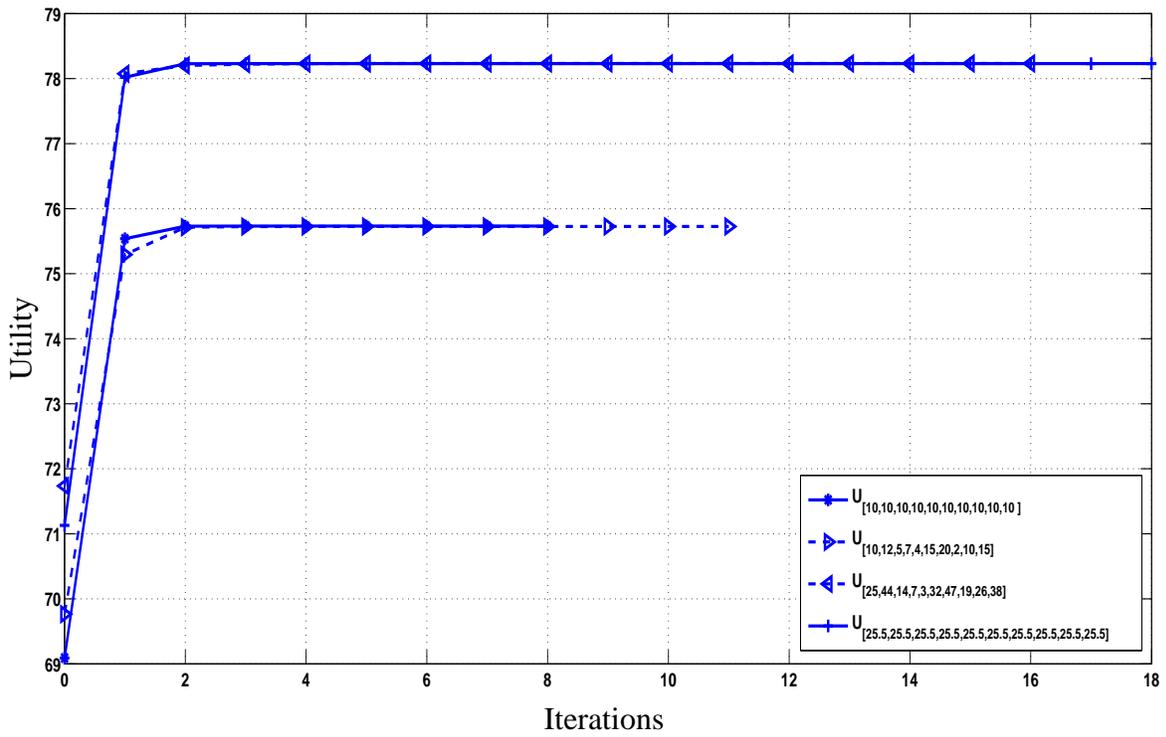

Fig. 4. Utility vs. Iterations (N=5, K=10): Starting from SG+TDMA, BCD converges to the optimal utility in 11,8,16,18 iterations for the following slot length sequences respectively: $S_1$, $\widetilde{S_1}$, $S_2$, $\widetilde{S_2}$.

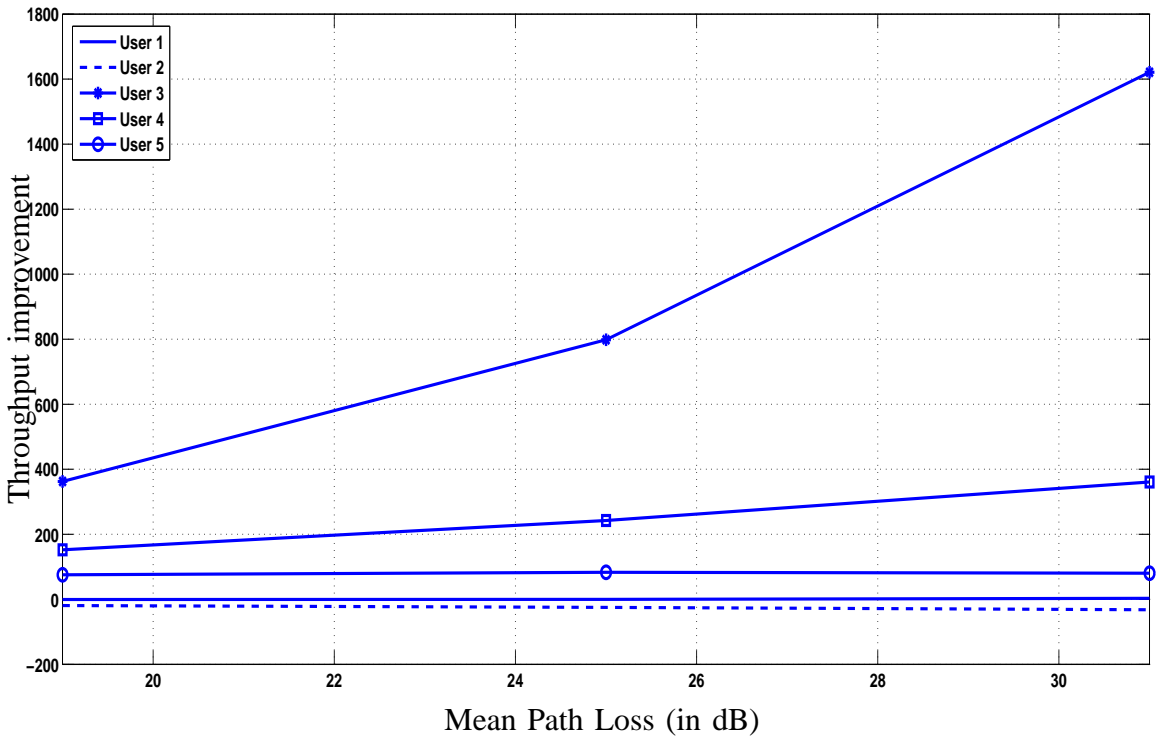

g replacements

Fig. 5.   Throughput Improvement vs. Mean Path Loss (N=5, K=10): Mean path loss is computed as the mean of the path losses of all users in the system. Results represent the throughput improvement of five users for three different path loss patterns. With minor decrease in the throughput of the stronger users, the weak users receive much more bits than that they used to receive with SG+TDMA.

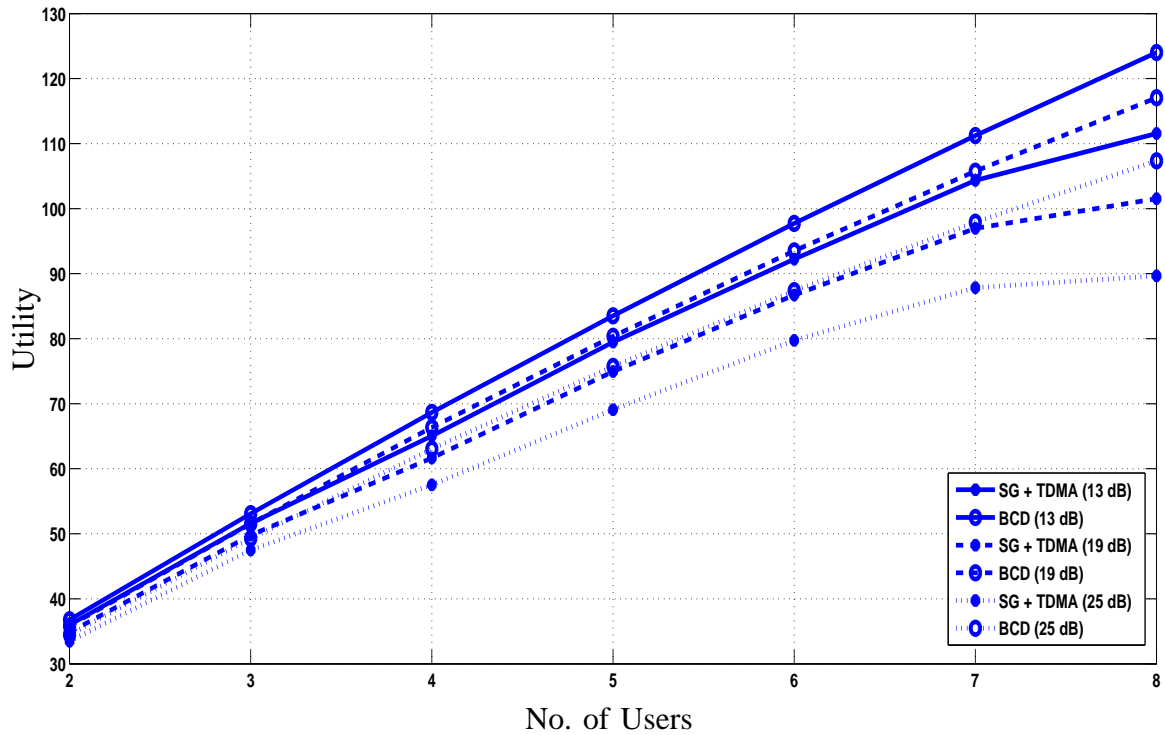

Fig. 6. Utility (SG+TDMA, BCD) vs. No. of Users: The utilities obtained by SG+TDMA and the proposed algorithm, for increasing number of users, are compared. The effect of path loss (the strongest user's path loss is shown between parentheses) on utility is shown. As path losses of the users increase the utility decreases.

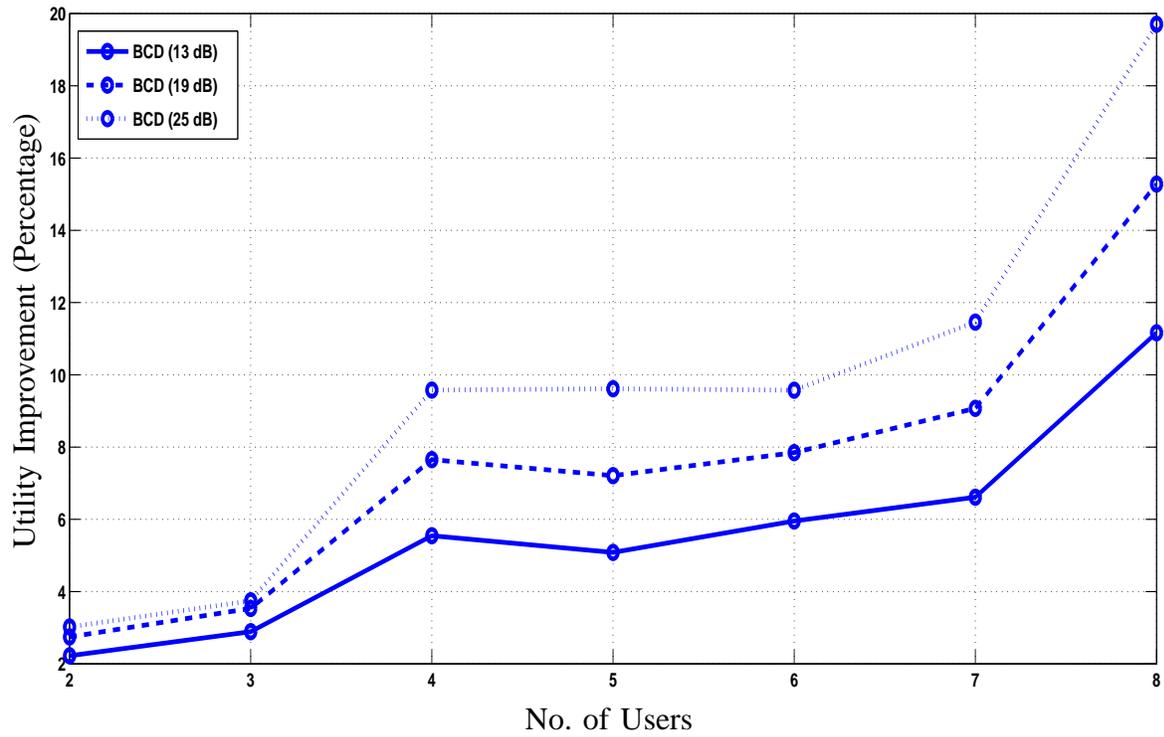

g replacements

Fig. 7. Utility Improvement vs. No. of Users: The utility improvement of the proposed algorithm over SG+TDMA, for increasing number of users, are compared. The effect of path loss (the strongest user's path loss is shown between parentheses) on utility improvement is shown. As path losses of the users increase the utility improvement increases, i.e, BCD's performance improves as the channel quality becomes degraded.

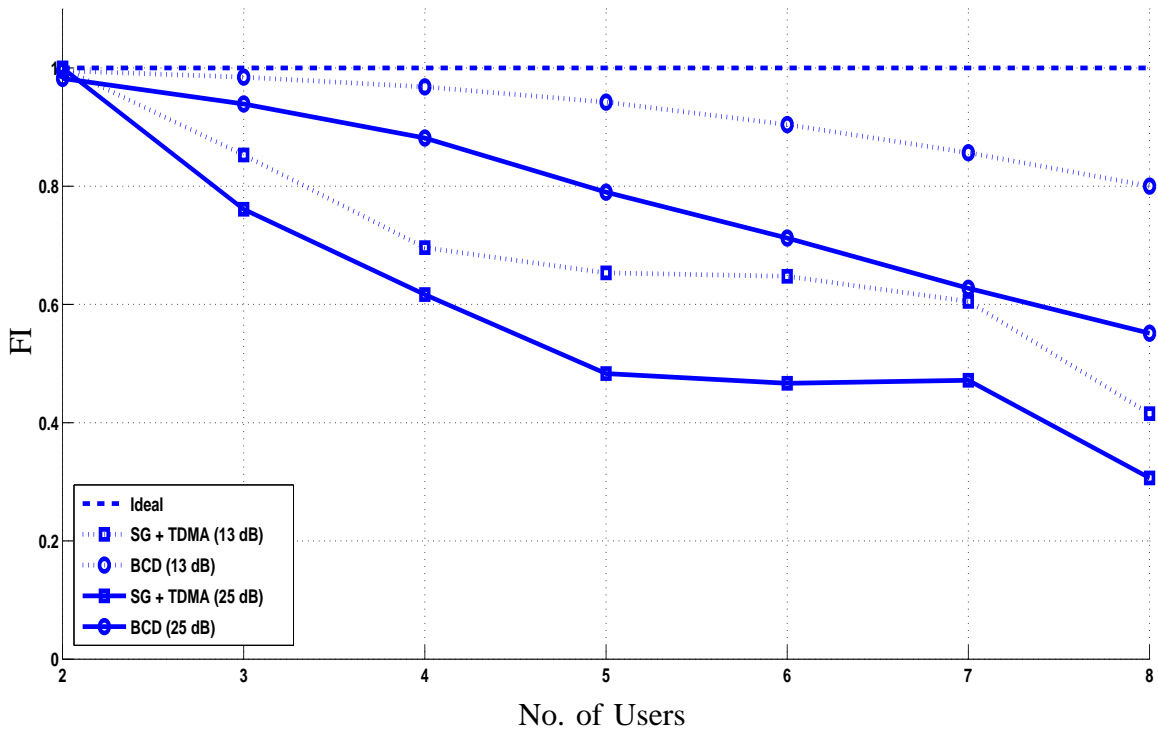

Fig. 8. Fairness Index (SG+TDMA, BCD) vs. No. of Users: The fairness of SG+TDMA and the proposed algorithm, for increasing number of users, are compared through $FI$, which takes the value of 1 when there is a complete fair allocation. The effect of path loss (the strongest user's path loss is shown between parentheses) on fairness is shown. As difference among the path losses of the users increase the fairness indexes of the schemes decrease (SG+TDMA being the worst), causing comparatively unfair allocations among users.